\documentclass[runningheads]{llncs}
\usepackage{graphicx,amsmath}
\usepackage{amsfonts}
\usepackage[]{algorithm2e}
\usepackage{multirow}
\usepackage{xcolor}
\PassOptionsToPackage{hyphens}{url}\usepackage{hyperref}
\begin{document}
\title{Weakly supervised multiple instance learning histopathological tumor segmentation}
\titlerunning{Weakly supervised multiple instance learning histopathological segmentation}
%
\author{Marvin Lerousseau\inst{1,2} \and
Maria Vakalopoulou\inst{1,2} \and
Marion Classe\inst{1,3} \and
Julien Adam\inst{3} \and
Enzo Battistella\inst{1,2} \and
Alexandre Carr\'e\inst{1} \and
Th\'eo Estienne\inst{1,2} \and
Th\'eophraste Henry\inst{1} \and
Eric Deutsch\inst{1} \and
Nikos Paragios\inst{4}}

%
\authorrunning{M. Lerousseau et al.}
%
\institute{Paris-Saclay University, Gustave Roussy, Inserm, 94800, Villejuif, France \and
Paris-Saclay University, CentraleSupélec, 91190, Gif-sur-Yvette, France \and
Gustave Roussy, Pathology Department, 94800, Villejuif, France \and TheraPanacea, 75014, Paris, France}
\maketitle              

\begin{abstract}
Histopathological image segmentation is a challenging and important topic in medical imaging with tremendous potential impact in clinical practice. 
State of the art methods rely on hand-crafted annotations which hinder clinical translation since histology suffers from significant variations between cancer phenotypes. 
In this paper, we propose a weakly supervised framework for whole slide imaging segmentation that relies on standard clinical annotations, available in most medical systems. 
In particular, we exploit a multiple instance learning scheme for training models.
The proposed framework has been evaluated on multi-locations and multi-centric public data from The Cancer Genome Atlas and the PatchCamelyon dataset. 
Promising results when compared with experts' annotations demonstrate the potentials of the presented approach. 
The complete framework, including $6481$ generated tumor maps and data processing, is available at \url{https://github.com/marvinler/tcga\_segmentation}.
%


\end{abstract}

\section{Introduction}
In digital pathology, whole slide images (WSI) are considered the golden standard for primary diagnosis~\cite{stathonikos2013going,jara2010digital}. The use of computer-assisted image analysis tools is becoming a mainstream for automatic quantitative or semi-quantitative analysis for pathologists, including the discovery of novel predictive biomarkers~\cite{janowczyk2016deep}. 
However, a lot of challenges remain to be addressed for machine learning methods because to the high variability of quality of tissue preparation and digital acquisition, and in tissue phenotype. 
A central objective in digital pathology is the accurate identification of cell or tissue of interest. 
For instance, computational staining of tumor tissue could be used for slide screening in order to increase the efficiency of pathologists. Automatically computed tumor maps could identify regions of interest for whole slide image classification~\cite{campanella2018terabyte}, or be combined with automatic detection of lymphocytes~\cite{saltz2018spatial} to further characterize the tumor and immune micro-environment for predicting treatment response~\cite{binnewies2018understanding}.

Traditionally, image segmentation is tackled by leveraging pixel-wise or patch-wise ground-truth annotations~\cite{guo2018review}. 
This is highly problematic in digital pathology due to the colossal size of WSIs with respect to the biological components, implying that the annotation process is considerably time-consuming.
Moreover, the high variance of clinical samples contributes on the deficiency of generalization, as illustrated in~\cite{campanella2019clinical} where the front-runner solution of the CAMELYON16 challenge~\cite{bejnordi2017diagnostic} has reportedly $4$ times higher classification errors on the same task for in-house data from the same location.

A standard multiple instance learning (MIL) scheme~\cite{dietterich1997solving} deals with classifying an entity (bag) from its constituents (instances). The MIL paradigm is particularly suited to histopathological image analysis due to its ability on reasoning on subsets of data (patches) that is often a computational necessity in histopathology. 
The general approach of MIL consists in learning a model that embeds each instance of a bag into a latent space. 
Then, the collection (usually of fixed size) of instances latent vectors is forwarded into an aggregating function which outputs the predicted bag probability, using different principles such as max-pooling~\cite{dietterich1997solving}, support vector machine~\cite{andrews2003support}, or even attention-based neural networks~\cite{ilse2018attention}. Recent large-scale histopathological studies provides promising classification solutions based on the MIL scheme~\cite{coudray2018classification,campanella2018terabyte,campanella2019clinical}. 
 Such approaches generally indicate whether a slide is non-neoplastic (normal), or the predicted subtype of apparent tumor tissue without accurately identifying the tumoral regions in the slide.

There are two ways to interpret multiple instance learning: MIL for classifying bags (or slides), or MIL for training an instance classifier model, apparent to bag segmentation. 
In particular, studies such as ~\cite{coudray2018classification,campanella2019clinical,campanella2018terabyte} use max-pooling MIL and its relaxed formulation~\cite{zhu2017deep} to first train an instance model, and then investigate various ways to combine instance predictions into a slide prediction.
These works demonstrate that MIL schemes provide powerful formulations for the WSI classification, by reaching AUC for tumor versus normal slide classification higher than $0.99$. 
However, these studies lack extensive evaluation for a more detailed MIL-driven segmentation performance since slide-based classification measures could lead to erroneous assessment regarding instance-level performance.

In this paper, we propose a weakly supervised segmentation scheme that is able to generate tumor segmentation models using annotations from the conventional clinical practice of pathologists' assessment. The contributions of this paper are: (i) a generic meta-algorithm, generating labels from WSI binary values intended to train detailed WSI segmentation models, (ii) a training scheme providing instance-level predictions, trained only with binary WSI annotations,  (iii) the release of 6481 automatically generated tumor maps for publicly available slides of The Cancer Genome Atlas (TCGA), an order of magnitude above previous released WSI tumor maps.


\section{Weakly supervised learning for tissue-type segmentation in histopathological images}
Contextually, we consider a set $S= \left\{S_i\right\}$ of training whole slide images, where each slide $S_i$ is associated with a label $T_i= \{0,1\}$ where $1$ refers to tumor and $0$ to normal.
 More precisely, $T_i=0$ indicates that there is no apparent tissue in the slide, and $T_i=1$ indicates that some tissue is tumorous. 
 The goal is to learn a tumor segmentation model, or a patch classifier, using only those binary annotations. 
To train a model in a fully supervised setup, a batch of patches $\left\{p_s\right\}$ is randomly sampled from a WSI along with their annotations computed beforehand.
However, such microscopic annotations are impractical to obtain when the number of whole slides images is in the hundreds or thousands, which is necessary for good generalization.
To deal with this limitation, the aim of the proposed framework is to generate a set of proxy patch-level ground truth labels by exploiting properties from the available global $T_i$ labels. 

By construction, a WSI with $T_i=0$ indicates that all extracted patches are of normal class ($0$). In that case, a proxy-label filled with $0$ provides perfect instance annotations, equivalent to a fully supervised learning scheme. However, in slides with $T_i=1$, tumor tissue could possibly be in any extent on the $S_i$. Alternatively, a WSI with $T_i=1$, normal tissue can theoretically cover no pixel up to the entire region in the slide except one pixel.
We integrate this property by proposing a training scheme in which two parameters $\alpha$ and $\beta$ are used for each training slide of $T_i=1$ in the following deterministic process:
\begin{itemize}
    \item assign a label $1$ to the patches ensuring at least $\alpha\%$ are of class $1$ 
    \item assign a label $0$ to the patches ensuring at least $\beta\%$ are of class $0$ 
    \item discard other patches from the computation of the loss signal
\end{itemize}
In such a setup, $\alpha\%$ represents the minimum assumed relative area of tumor tissue in the WSI, and similarly for  $\beta\%$ with the normal tissue extent. Because the explicit process is deterministic, the framework is identified by the values of $\alpha$ and $\beta$. Noteworthy, $(\alpha, \beta)$ such that $(\alpha+\beta)>100\%$ would produce contradictory proxy labels for $100-(\alpha+\beta)$\%$>0$ instances, which could impede training by diminishing the signal-to-noise ratio. Therefore, the possible space of these two parameters can be defined as $\mathbb{F}=\{(\alpha, \beta); \alpha>0, \beta\geq0, \alpha+\beta\leq1\}$. 

Formally, given a loss function $L$ (e.g. binary cross-entropy), the formulated framework aims at minimizing the following empirical risk on $S$:
\begin{align*}
    c_0\cdot \sum_{\substack{S_i \in S;\\ T_i=0}}&\bigg[\underbrace{\sum_{p_s\mathtt{\sim}S_i} L(f(p_s), 0)}_{R_\text{FP}}\bigg]\\
    + &c_1\cdot\sum_{\substack{S_i \in S;\\ T_i=1}}\bigg[\underbrace{\sum_{\substack{p_s\mathtt{\sim}S_i;\\ p_s \in P(f(p_s); \alpha, 100)}} L(f(p_s), 1)}_{R_\alpha}+\underbrace{\sum_{\substack{p_s\mathtt{\sim}S_i;\\ p_s \in P(f(p_s); 0, \beta)}} L(f(p_s), 0)}_{R_\beta}\bigg]
\end{align*}
where $P(f(p_s); p_\text{min}, p_\text{max})$ is defined as the subset of patches ${f(p_s)}$ for which the predicted probability lies within the $p_\text{min}$\textit{th} and the $p_\text{max}$\textit{th} percentiles of the predictions $f(p_s)$, $c_0$ and $c_1$ are constants for batch averaging and class imbalance for both classes, and $T_i$ refers to the binary ground-truth of $S_i$. 
Minimizing this empirical risk will guide models into recalling enough positive tumoral patches ($R_\alpha$) per slide but not too much ($R_\beta$) while maintaining a low level of false positive in negative slides $R_\text{FP}$.
The formulated approach is generic, in the sense that it can be used to train a large scope of machine learning models, including neural networks, with patch-based or pixel-wise predictions, and it can be coupled with most usual loss functions. It produces trained segmentation models, readily available to produce heatmaps without intervention of the formulated pipeline nor $\alpha\%$ and $\beta\%$ parameters. 

 \begin{figure}[t!]
\includegraphics[width=\textwidth]{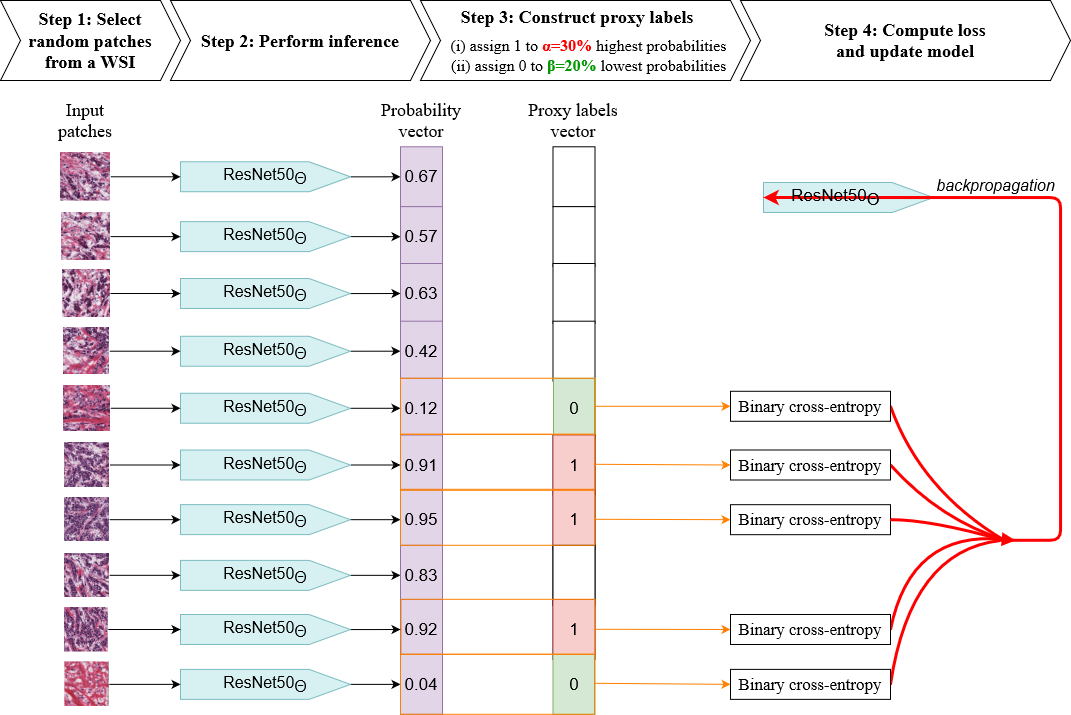}
\caption{Illustration of the processing of a batch of $10$ patches from a positive WSI. A unique ResNet50 model with parameters $\theta$ is inferred on all images of the batch. For a given configuration of $(\alpha, \beta)$ (here, $(0.3, 0.2)$), these predictions are first used to create a proxy image-wise labels vector, then combined with the proxy label to compute batch error for further parameters $\theta$ update with backpropagation.}
\label{fig:pipeline}
\end{figure}

\section{Implementation details and Dataset}

\subsection{Framework setup and Architecture details}
We perform a benchmark of a representative population of the framework parameters space  $\mathbb{F}$. Specifically, $\mathbb{F}$ is sampled starting from 0 with increment of $0.2$ (or 20\%) for both $\alpha$ and $\beta$ (\textit{e.g.} $(0, 0)$, $(0, 0.2)$, $(0, 0.4)$ and so on), resulting in $\frac{6\cdot 7}{2}=21$ configurations. Of those, the $6$ configurations with $\alpha=0$ are discarded, as this would imply that the framework provides only $0$ labels contradicting with the $T_i=1$ assumption of our empirical risk formulation. At the end only $15$ sampled configurations had been used.

Each configuration is used to train a ResNet50 architecture~\cite{he2016deep}, which has been extensively used for histopathology image analysis in a multitude of tasks~\cite{mormont2018comparison}, and can be used without the global average pooling layer to yield $13\times 13$ outputs per $224$ pixel-wide input image. Pre-training is used with initialization on a well-performing snapshot on ImageNet~\cite{deng2009imagenet}. At each epoch, each training slide is sampled once. Upon sampling, a batch size of 150 patches of size $224\times 224$ are randomly sampled at $20$x magnification in the tissue region of a WSI.
Data augmentation is used independently on each image, with random rotations and flips while also applying channel-wise standard scaling from training averages and variances, and color jitter.
The model is then concurrently inferred on the $150$ patches of the batch, and a proxy-vector is constructed with the formulated pipeline as illustrated for $10$ patches in Figure~\ref{fig:pipeline}. Specifically, the $\lfloor150\times\alpha\rfloor$ patches of highest probabilities are associated with a proxy value of 1, and the $\lfloor150\times\beta\rfloor$ patches of lowest probabilities with a proxy value of 0. A masking vector of size $150$ is concurrently filled such that patches with no attributed label are discarded. The proxy vector is then coupled with the model's $150$ predictions minus the discarded ones, in order to compute patch-wise binary cross-entropy loss which is then averaged and retro-propagated across all the non masked predictions.
  The error signal is used to tune model parameters using Adam optimizer with learning rate of $10^{-4}$ and default momentum parameters. $c_0$ and $c_1$ are set to $1$. Each configuration is trained for 20 epochs on $2$ V100 NVIDIA GPU, for a training time of  $\mathtt{\sim}16$hours, or a total benchmark training time of $\mathtt{\sim}240$hours. The code is implemented with pytorch 1.5.0 and torchvision 0.4.0 on python3.6.

\subsection{Dataset}
The dataset consists of 6481 flash-frozen whole slide images from TCGA, issued from kidney (2334), bronchus and lung (2168) and breast (1979) WSIs locations. 
These locations were selected on TCGA as the first $3$ indexed, while no slide filtering nor slide selection has been performed to be coherent with standard clinical practices. This dataset was divided in training, validation and testing sets on a case basis, with $65\%$, $15\%$, and $20\%$ of cases respectively. For the rest of the paper this testing set is denoted as ''In-distribution''. 
Each selected configuration is trained using the training set, with hyper-parameters optimized on the validation set. Then, their performance is assessed on the testing set. For extensive quantitative performance assessment, expert pathologists annotated 130 slides from this testing set (45 breast, 40 kidney, 45 bronchus and lung), thus measuring in-distribution generalization. Annotations were computed at 20x magnification by a junior pathologist on a in-house annotation tool by contouring tumor tissue regions which were then filled, and were modified until validation by a senior pathologist. 

The same protocol was applied on additional slides extracted from locations which are not used in the previous cohort. Specifically, 35 WSIs from colon, 35 from ovary and 30 from corpus uteri are pixel-wise annotated and used to measure generalization performance of models to unseen tissue environments, which we denote as the "Out-of-location" testing set. 
 We pinpoint that these annotations were not used during training nor validation, but only to assess testing segmentation performance of the produced models. For training, we use diagnostic labels extracted directly from TCGA, for which each slide name contain a code indicating the type of specimen (e.g. "Solid Tissue Normal" or "Primary Solid Tumor")\footnote{https://gdc.cancer.gov/resources-tcga-users/tcga-code-tables/sample-type-codes}. Notably, normal slides are explicitly discerned from slide with apparent pathological tissue. In such context, each slide is associated with a binary value indicating whether tumor tissue is apparent in the slide, or whether the slide is of normal tissue only. To further compare with results from the community, we infer all models on the PatchCamelyon dataset~\cite{veeling2018rotation}. The dataset consists in $96\times 96$ patches extracted from formalin-fixed, paraffin-embedded (FFPE) tissues from sentinel lymph nodes at 10x magnification. In PatchCamelyon, images are labeled as $1$ if their $32\times 32$ center region contains at least 1 pixel of tumor cell, otherwise $0$. To accommodate with the $224\times 224$ input at 20x magnification of the learned models, these images were bi-linearly upsampled twice and padded with 0. This testing set is particularly challenging for the benchmarked models, since they are not trained on FFPE slides, which  are visually highly different from flash-frozen ones. Besides, the trained models did not include any tissue extracted from sentinel lymph nodes, highlighting the generalization challenge of the proposed framework.

\section{Results and Discussion}
\begin{figure}[t!]
\includegraphics[width=\textwidth]{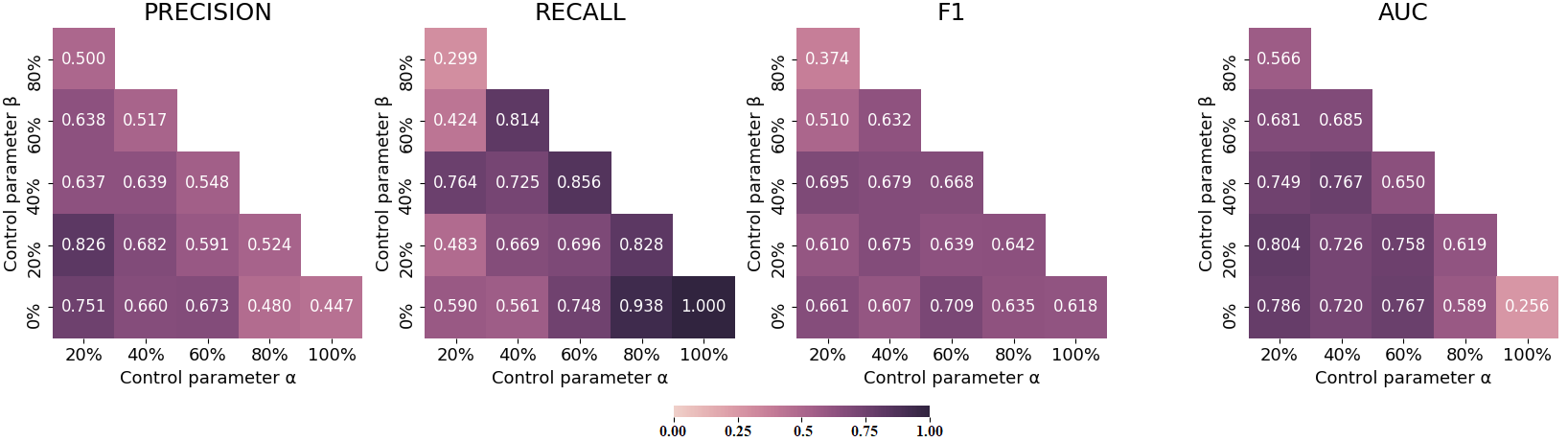}
\caption{Quantitative results for the 15 benchmarked configurations on the hold-out testing set (In-distribution) from bronchus and lung, kidney, and breast locations. Each subplot (4 in total) displays a pixelwise measure, as indicated in its sup-title, for each configuration in a matrix format. AUC: area under the ROC curve.}
\label{fig:quantitative}
\end{figure}

For performance assessment, all $15$ trained ResNet50 models are inferred on the testing slides. The resulting heatmaps are compared to segmentations maps provided by the pathologists.
All configurations are found to converge to sup-random In-distribution performance, except for the two extreme configurations $(\alpha=1, \beta=0)$ and $(0.2, 0.8)$, as displayed in Table~\ref{tab:auc_pixel}. The average In-distribution AUC is $0.675 \pm 0.132$, with optimal AUC of $0.804$ for $(\alpha=0.2, \beta=0.2)$. Precision and recall are extracted after threshold selection on validation set and displayed in Figure~\ref{fig:quantitative}. The $\alpha$ parameter seems to influence the recall at the  the  expense  of precision. Upon performance introspection by location, all configurations report the worse performance for the location bronchus and lung, with twice as much AUC error compared to kidney and breast locations. Concerning the Out-of-distribution cohort, the average AUC is $0.679 \pm 0.154$, which is close to In-distribution performance, although lower when omitting bronchus and lung location from the latter. There is no evident pattern for configurations that yield improved Out-of-location results.

\begin{table}
\caption{Pixelwise AUC for the 15 $(\alpha, \beta)$ framework configurations on the hold-out testing set (In-distribution) and the testing set from locations unseen in training (Out-of-location). Grey results take background into account, black ones are computed by completely discarding background from performance computation.}
\begin{tabular}{c|lllll|llll|lll|ll|l}
\multicolumn{1}{r|}{$\alpha$ =} & \multicolumn{5}{c|}{0.2}                        & \multicolumn{4}{c|}{0.4}                & \multicolumn{3}{c|}{0.6}        & \multicolumn{2}{c|}{0.8} & \multicolumn{1}{c}{1.0}   \\
\multicolumn{1}{r|}{$\beta$ =}  & \multicolumn{1}{c}{0} & \multicolumn{1}{c}{0.2} & \multicolumn{1}{c}{0.4} & \multicolumn{1}{c}{0.6} & \multicolumn{1}{c|}{0.8} & \multicolumn{1}{c}{0} & \multicolumn{1}{c}{0.2} & \multicolumn{1}{c}{0.4} & \multicolumn{1}{c|}{0.6} & \multicolumn{1}{c}{0} & \multicolumn{1}{c}{0.2} & \multicolumn{1}{c|}{0.4} & \multicolumn{1}{c}{0} & \multicolumn{1}{c|}{0.2} & \multicolumn{1}{c}{0} \\ \hline
                                   & .786                        & .804                        & .749                        & .681                        & .566                        & .720                        & .726                        & .767                        & .685                        & .766                        & .758                        & .650                        & .589                        & .619                        & .256                        \\
\multirow{-2}{*}{In-distribution}  & {\color[HTML]{656565} .964} & {\color[HTML]{656565} .952} & {\color[HTML]{656565} .960} & {\color[HTML]{656565} .930} & {\color[HTML]{656565} .874} & {\color[HTML]{656565} .967} & {\color[HTML]{656565} .953} & {\color[HTML]{656565} .959} & {\color[HTML]{656565} .935} & {\color[HTML]{656565} .957} & {\color[HTML]{656565} .960} & {\color[HTML]{656565} .940} & {\color[HTML]{656565} .946} & {\color[HTML]{656565} .946} & {\color[HTML]{656565} .926} \\
\multicolumn{1}{l|}{}               &                              &                              &                              &                              &                              &                              &                              &                              &                              &                              &                              &                              &                              &                              &                              \\
                                   & .866                        & .732                        & .709                        & .583                        & .257                        & .783                        & .710                        & .762                        & .658                        & .790                        & .787                        & .673                        & .785                        & .695                        & .404                        \\
\multirow{-2}{*}{Out-of-location}  & {\color[HTML]{656565} .984} & {\color[HTML]{656565} .974} & {\color[HTML]{656565} .972} & {\color[HTML]{656565} .958} & {\color[HTML]{656565} .917} & {\color[HTML]{656565} .980} & {\color[HTML]{656565} .972} & {\color[HTML]{656565} .978} & {\color[HTML]{656565} .966} & {\color[HTML]{656565} .981} & {\color[HTML]{656565} .980} & {\color[HTML]{656565} .968} & {\color[HTML]{656565} .980} & {\color[HTML]{656565} .970} & {\color[HTML]{656565} .933}
\end{tabular}
\label{tab:auc_pixel}
\end{table}

\begin{figure}[t!]
\includegraphics[width=\textwidth]{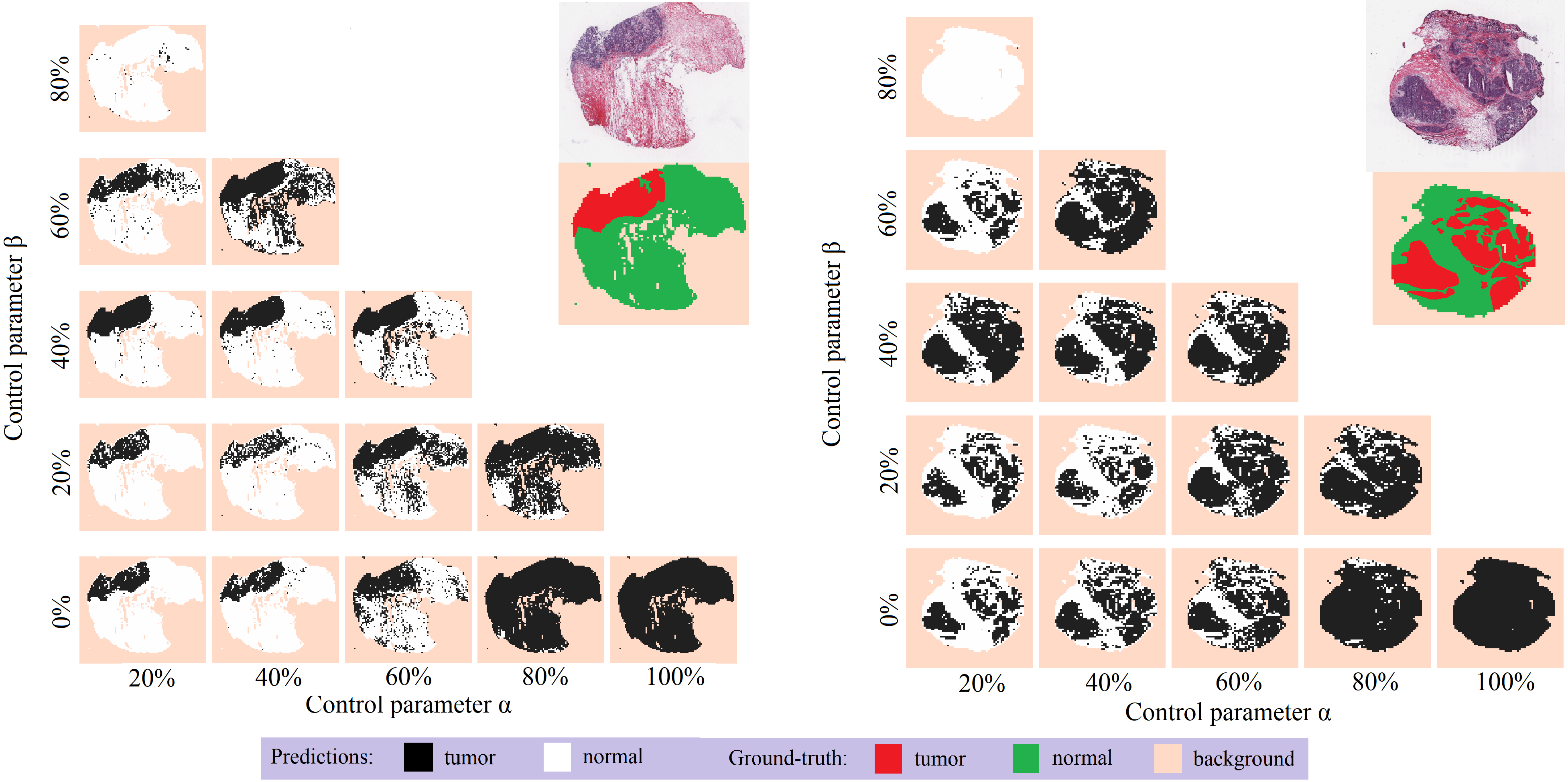}
\caption{Unfiltered predicted tumor maps on hold-out testing samples for the 15 benchmarked framework configurations. 2 WSI and their corresponding results are displayed in a matrix-format. The red and green images are the ground-truth.}
\label{fig:qualitative}
\end{figure}

Some visual representations of two different samples testing are presented on Figure~\ref{fig:qualitative}. In particular, in the figure we present the WSI image together with the pixel-wise annotations of the pathologist. Moreover, different segmentation maps depending on the configuration are also presented. It can observed that there are $3$ or $4$ configurations that are close to the expert's annotations. These configurations are in line with our quantitative results.  Additional post-processing strategies would potentially boost the performance of our framework.

To test the generalization of our method, we performed also experiments on the PatchCamelyon dataset~\cite{veeling2018rotation}. 
In particular we found the most of the configurations ($12$ out of the $15$) reporting quite low AUC, between $0.428$ and $0.612$.  However, $3$ configurations are found to generalize to some extent, that is $(\alpha=0.2, \beta=0)$ with $0.672$ AUC, $(0.2, 0.2)$ with $0.802$, and $(0.4, 0)$ with $0.758$. Although these results are far from report AUC of $0.963$ obtained with fully supervised models specifically trained on this dataset~\cite{veeling2018rotation}, the results suggest the presented framework could provide models which can grasp generic discriminative cancer features from multiple types of slides in broad biological context.

\section{Conclusion and future works}
In this paper we propose a weakly supervised model which provides segmentation maps of WSIs, trained only with binary annotations over the entire WSIs. From our experiments we saw that usually $3$ to $4$ configurations are expected to yield respectively high precision, high recall, and high overall performance for WSIs of different organs and tumor coverage. The findings in this paper highlight the potential of weakly supervised learning in histopathological image segmentation, which is known to be heavily impeded by the annotation bottleneck.
With the complete open-source releases of both the complete WSI pre-processing pipeline, the presented training framework, as well as the inference pipeline, the presented approach can be used off-the-shelf for pan-cancer tumor segmentation using the entire $18$k flash-frozen WSI of TCGA, or other type of tissue segmentation such as necrosis or stromal tissue. The public release of $6481$ automatically generated tumor maps, with an expected AUC above 0.932, should lower the barrier of entry to pathomics by bypassing tumor annotation efforts. All code and results can be found at \url{https://github.com/marvinler/tcga\_segmentation}.

There are many ways to fine-tune a segmentation model using the formulated framework, such as with more appropriate deep learning architectures or with more extensive hyper-parameters optimization. We believe the most impactful future works will revolve around the proxy-generation labels from more sophisticated slide labels which would yield higher information while remaining fast to obtain, essentially trading annotation time for performance.


\bibliographystyle{splncs04}
\bibliography{paper2160}
\end{document}